# Reaction of $C_2(a^3\Pi_u)$ with methanol: Temperature dependence and deuterium isotope effect

Renzhi Hu, Qun Zhang* and Yang Chen*

*Hefei National Laboratory for Physical Sciences at the Microscale and Department of Chemical Physics, University of Science and Technology of China, Hefei, Anhui 230026, P. R. China.*

*E-mail:* qunzh@ustc.edu.cn and yangchen@ustc.edu.cn

## ABSTRACT

Bimolecular rate constants for the gas-phase reactions of $C_2(a^3\Pi_u)$ with a variety of methanol isotopomers including $CH_3OH$ ($k_1$), $CH_3OD$ ($k_2$), $CD_3OH$ ($k_3$), and $CD_3OD$ ($k_4$) have been measured over the temperature range 293 − 673 K by means of pulsed laser photolysis/laser-induced fluorescence technique. The rate constants, in the units $cm^3$ $molecule^{-1}$ $s^{-1}$, can be fitted by the normal Arrhenius expressions: $k_1(T) = (1.32 \pm 0.02) \times 10^{-11} \exp[-(366.80 \pm 4.44)/T]$, $k_2(T) = (1.34 \pm 0.02) \times 10^{-11} \exp[-(376.86 \pm 5.09)/T]$, $k_3(T) = (1.09 \pm 0.02) \times 10^{-11} \exp[-(640.00 \pm 7.23)/T]$, and $k_4(T) = (1.12 \pm 0.01) \times 10^{-11} \exp[-(666.37 \pm 4.63)/T]$, where all error estimates are ±2σ and represent the precision of the fit. The observed deuterium kinetic isotope effects, $k_1/k_2$ and $k_1/k_3$, along with the positive temperature dependences of $k(T)$, allow us to reach a conclusion that the reaction of $C_2(a^3\Pi_u)$ with methanol in 293 – 673 K proceeds *via* a site-specific hydrogen abstraction mechanism, that is, H-atom abstraction from the methyl rather than "uninvolved" hydroxyl site dominates reactivity.

# Introduction

The dicarbon molecule, $C_2$, is believed to be present in a wide variety of environments such as hydrocarbon flames,[1] stellar atmospheres,[2,3] comets,[4] circumstellar envelopes,[5] interstellar clouds,[6,7] and plasmas.[8,9]  In addition to being one of the simplest diatomic molecules, $C_2$ is a rather unusual species as it possesses a low-lying first excited metastable triplet state ($a^3\Pi_u$) which is only ~610 cm$^{-1}$ (*i.e.*, ~1.7 kcal mol$^{-1}$ or ~880 K) above its ground singlet state ($X^1\Sigma_g^+$).[10]  $C_2$ in both states can be readily detected by means of laser-induced fluorescence technique *via* the well-known Swan ($d^3\Pi_g - a^3\Pi_u$) and Mulliken ($D^1\Sigma_u - X^1\Sigma_g$) bands,[11] which facilitates the kinetic studies of its reactions with other co-reagent species in various chemical systems.

Research on reaction kinetics of $C_2$ has surged up since the late seventies of last century, most of which concerned its reactions with such molecules as small hydrocarbons and inorganic species.[12-34]  While extensive kinetic data have been accumulated over the past three decades, very few has been reported relating to the $C_2$ reactions with oxygenated volatile organic compounds (OVOCs), a class of chemicals that is believed to play a crucial role in atmospheric chemistry.[35]  We recently reported on the room-temperature kinetics of $C_2(a^3\Pi_u)$ reactions with a series of alcohols (*i.e.*, methanol, ethanol, propanol, butanol, and pentanol),[29] an important class of OVOCs.  To the best of our knowledge, this is the only available report in this regard.

It was argued in the previous work[29] that the reactions of $C_2(a^3\Pi_u)$ with small alcohol molecules at room temperature proceeds *via* a mechanism of hydrogen abstraction, which was based on the bond dissociation energy (BDE)[36-38] and linear free energy[37] correlations together with an *ab initio* calculation performed on the reaction of $C_2(a^3\Pi_u)$ with $CH_3OH$. Noteworthy is that the BDE correlation of alcohols given in that work was performed only against corresponding alkanes.[29] Nevertheless, it has been well recognized that an H/D isotopomerization based BDE correlation can usually be more straightforward and reliable.[39-45]

We present here for the first time a detailed examination of deuterium kinetic isotope effect as well as temperature dependence for the reaction of $C_2(a^3\Pi_u) + \text{methanol}$, aiming to elucidate its reaction mechanism. Bimolecular rate constants for the reactions

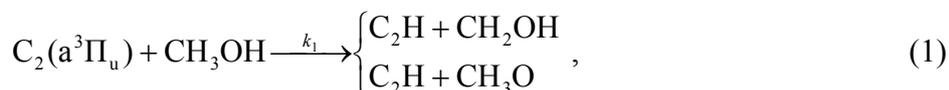

$$C_2(a^3\Pi_u) + CH_3OH \xrightarrow{k_1} \begin{cases} C_2H + CH_2OH \\ C_2H + CH_3O \end{cases}, \qquad (1)$$

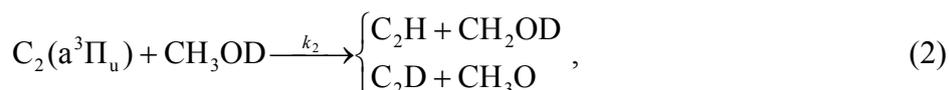

$$C_2(a^3\Pi_u) + CH_3OD \xrightarrow{k_2} \begin{cases} C_2H + CH_2OD \\ C_2D + CH_3O \end{cases}, \qquad (2)$$

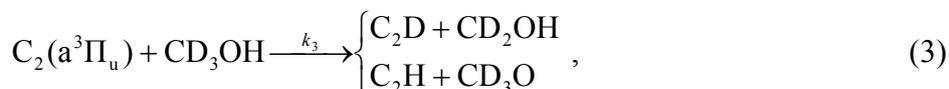

$$C_2(a^3\Pi_u) + CD_3OH \xrightarrow{k_3} \begin{cases} C_2D + CD_2OH \\ C_2H + CD_3O \end{cases}, \qquad (3)$$

and 

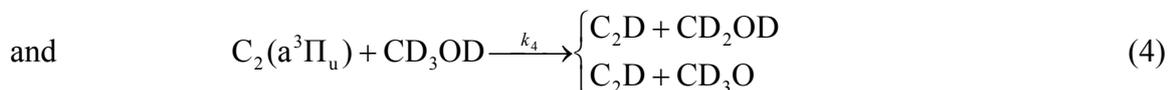

$$C_2(a^3\Pi_u) + CD_3OD \xrightarrow{k_4} \begin{cases} C_2D + CD_2OD \\ C_2D + CD_3O \end{cases} \qquad (4)$$

were measured over the temperature range 293 − 673 K using the conventional pulsed laser photolysis/laser-induced fluorescence method. The observed kinetic isotope effects, $k_1/k_2$ and $k_1/k_3$, along with their temperature dependences are put in detail.

## Experimental

The experiment was performed under slow-flow conditions by means of pulsed laser photolysis/laser induced fluorescence technique described in detail elsewhere.[26-30] Briefly, the $C_2(a^3\Pi_u)$ radicals were produced by photolysis of $C_2Cl_4$ using the fourth harmonic output (266 nm, typical energy ~5.5 mJ/pulse, repetition rate 10 Hz) of a Nd:YAG laser (INDI, Spectra Physics). An $f$ = 500 mm lens was used to focus the photolysis laser beam into the center of the stainless steel flow reactor. The probe laser used was a Nd:YAG laser (GCR-170, Spectra Physics) pumped dye laser (PrecisionScan, Sirah) whose output energy was chosen to be ~1.5 mJ/pulse. The dye laser beam was set collinear to the photolysis laser beam in a counter-propagating way and its diameter was confined to ~1 mm in order to effectively suppress the scattered light.

The $C_2(a^3\Pi_u)$ concentrations were probed by exciting the $d^3\Pi_g \leftarrow a^3\Pi_u$ 0-0 transition at 516.5 nm[11] and detecting the dye laser induced fluorescence (LIF) on the 0-1 vibronic transition band which was isolated with an interference filter at 563.5 nm. The LIF signal was collected by a photomultiplier tube (R928, Hamamatsu) whose output was recorded by a digital oscilloscope (TDS380, Tektronix) and then averaged over 256 laser pulses with a computerized data acquisition system. A digital delay generator (DG535, Stanford Research) was used to vary the time delay between the photolysis and probe laser pulses.

Temperatures of the resistively heated flow reactor were controlled over the range 293 − 673 K by a temperature regulator (AI-708P, Xiamen) and measured by a K-type thermocouple probe mounted a few millimeters away from the center of the reactor. The controlled temperatures were found to be within a ±0.5 K precision over the dimensions of the probed volume and the experimental duration.

In a typical experiment, three channels of gas flow including premixed $C_2Cl_4$/Ar, methanol/Ar, and pure argon gas slowly passed through the reaction cell and were measured independently by three calibrated mass flow controllers (D07-7A/2M, Beijing). The concentrations of methanol reactants can be calculated by

$$[\text{methanol}] = 9.66 \times 10^{18} \frac{p}{T} \frac{f_{\text{methanol/Ar}} x_{\text{methanol/Ar}}}{f_{\text{total}}}, \quad (5)$$

where [methanol] is the concentration of methanol reactant (in molecules cm$^{-3}$), $p$ and $T$ are the pressure (Torr) and temperature (K) of the system, $f_{\text{methanol/Ar}}$ and $f_{\text{total}}$ are the flow rates of methanol/Ar and all gases, respectively, and $x_{\text{methanol/Ar}}$ is the pressure ratio of methanol over Ar. All the experiment was performed over the temperature range 293 − 673K and at a total pressure of ~9.5 Torr to maintain a steady flow condition.

The $C_2(a^3\Pi_u)$ radicals generated from the laser photolysis of the precursor $C_2Cl_4$ molecules are rotationally hot. The Ar buffer gas was included in the reaction mixtures in order to cool $C_2(a^3\Pi_u)$ radicals rotationally and translationally as well as to slow down the diffusion of molecules out of the detection region.[18] The LIF decay of $C_2(a^3\Pi_u)$ was measured after $C_2(a^3\Pi_u)$ is rotationally cooled to room temperature. Under the above experimental conditions, since the $C_2(a^3\Pi_u)$ radicals were found to be rotationally cooled to room temperature within a delay time of ~8 μs between the photolysis and probe laser pulses,[26-30] the kinetic data were collected at times slightly longer than 8 μs.

The chemicals used in the experiment were as follows: $C_2Cl_4$ (≥98.5%, Shanghai), methanol (≥99.5%, Shanghai), methanol-$d$ (isotopic purity 99.0% at% D, Acros), methanol-$d_3$ (99.5% at% D, Acros), methanol-$d_4$ (99.8% at% D, Aldrich). All samples were degassed by repeated freeze-pump-thaw cycles in liquid nitrogen. Ar (99.999%, Nanjing gas) was used without further purification.

## Results and discussion

In the present study, the UV laser photolysis of $C_2Cl_4$ produces $C_2$ radicals in both $X^1\Sigma_g^+$ and $a^3\Pi_u$ states. Since the singlet and triplet states of $C_2$ are quite close in energy,[10] population transfer between the two states through collisions either with the carrier gas (Ar) or with the co-reagent methanol molecules can occur *via* intersystem crossing.[18,20,33] Although such a collision-induced quenching process cannot be distinguished from reaction in the present work, it can be justified to be negligible compared to reaction based on the following considerations. On the one hand, an upper limit of rate coefficient for the $a^3\Pi_u \rightarrow X^1\Sigma_g^+$ transfer by Ar has been determined to be $3 \times 10^{-14}$ $cm^3$ $molecule^{-1}$ $s^{-1}$,[18] which is roughly two orders of magnitude smaller than that obtained for reactions in this work. On the other hand, as a singlet quencher consisting of light atoms, methanol molecule cannot induce significant intersystem crossing due to the fact that spin conservation rules work rather well for light species.[18,20] In addition, $C_2(a^3\Pi_u)$ was found to be virtually produced in its vibrational levels up to $\upsilon = 3$ (as verified by the LIF excitation spectrum of the $C_2$ Swan bands we obtained in the wavelength range 506 – 518 nm). The buffer gas Ar used in this experiment ensures rapid rotational and translational equilibration, however it cannot efficiently quench vibrational excitation as He or $N_2$ does.[18,20] In contrast, $CH_4$ was found to be rather efficient at removing $\upsilon = 1$ population.[13] Nevertheless, with the exception of the $C_2 + O_2$ reaction, rate coefficients were demonstrated to be identical using $N_2$ and $CH_4$ buffer gases, ruling out any artifacts in the rate data due to filling of $\upsilon = 0$ from upper levels.[13] As in this case, although only the disappearance of $\upsilon = 0$ was monitored in the present work, the vibrationally hot $C_2(a^3\Pi_u)$ does not complicate the interpretation of the measured rates.

Similar to our previous work,[26-30] the validity of pseudo-first-order kinetics was ensured throughout all the measurements presented here by keeping the partial pressure of the $C_2Cl_4$ precursor always much lower than that of the methanol reactants. Under such a condition, the LIF decay of $C_2(a^3\Pi_u)$ at long times (>8 μs) follows

$$I = I_0 \exp(-k't), \qquad (6)$$

where $k'$ is the pseudo-first-order rate constant for the total loss of the $C_2(a^3\Pi_u)$ LIF signal $I$ due to all processes including reactions and diffusion out of the probe region, and $t$ is the time delay between the photolysis and probe lasers. Fig. 1 shows a typical time-resolved LIF decay profile for the co-reagent species $CH_3OD$ with a concentration of $1.42 \times 10^{15}$ molecules $cm^{-3}$ at 573 K. By varying reactant concentrations we obtained the corresponding $k'$ values through exponential fits using eqn (6).

As a result of the pseudo-first-order kinetics, $k'$ at a certain concentration of methanol under investigation is given by

$$k' = k[\text{methanol}] + k_n, \qquad (7)$$

where $k$ is the bimolecular rate constant for the reaction of $C_2(a^3\Pi_u)$ with methanol, and $k_n$ is the loss rate of $C_2(a^3\Pi_u)$ due to the reactions and diffusion in the absence of methanol. Linear least-squares fit of $k'$ versus [methanol] therefore yields the bimolecular rate constant $k$ for the reaction of $C_2(a^3\Pi_u)$ with methanol. Fig. 2, as an example, shows plots of $k'$ as a function of the concentrations of $CH_3OH$, $CD_3OH$, $CH_3OD$, and $CD_3OD$ at 373, 373, 473 and 473 K, respectively, all of which exhibit excellent linear relationship.

Table 1 lists the obtained bimolecular rate constants (in the units $10^{-12}$ $cm^3$ molecule$^{-1}$ s$^{-1}$, within ±2σ errors) for the reactions of $C_2(a^3\Pi_u)$ with four isotopomers of methanol including

CH$_3$OH ($k_1$), CH$_3$OD ($k_2$), CD$_3$OH ($k_3$), and CD$_3$OD ($k_4$) over the temperature range 293 − 673 K.

The rate constants measured over the temperature range 293 – 673 K for reaction (1), $k_1(T)$, are plotted in Fig. 3. Also included in Fig. 3, for comparison purposes, are the previously reported data measured at room temperature (shown as the open diamond),[29] with which the present room-temperature measurement is in reasonably good agreement. The fit of the temperature dependent $k_1$ data (in the units cm$^3$ molecule$^{-1}$ s$^{-1}$) to Arrhenius equation of the normal form $k_1(T) = A \exp(-E/RT)$ yields

$$k_1(T) = (1.32 \pm 0.02) \times 10^{-11} \exp[-(366.80 \pm 4.44)/T], \qquad (8)$$

which shows clearly an excellent linear relationship on the semi-logarithm scale. The appearance of a positive temperature dependence (*i.e.*, a positive activation energy as shown in eqn (8)) indicates that there exists an energy barrier along the reaction coordinate, which in turn implies that a hydrogen abstraction instead of an insertion mechanism most probably takes place in the course of the reaction.[20]

In an attempt to gain more insights into the reaction mechanism, we further examined the kinetic isotope effects by replacing CH$_3$OH with deuterated CH$_3$OD, CD$_3$OH, and CD$_3$OD.

The rate constants measured over the temperature range 293 – 673 K for reaction (2), $k_2(T)$, are also plotted in Fig. 3. The Arrhenius fit of $k_2$ data (in the units cm$^3$ molecule$^{-1}$ s$^{-1}$) reads

$$k_2(T) = (1.34 \pm 0.02) \times 10^{-11} \exp[-(376.86 \pm 5.09)/T]. \qquad (9)$$

As can be readily seen from Fig. 3, $k_2$ exhibits a positive temperature dependence almost identical to that of $k_1$ and $k_2(T) \approx k_1(T)$ at each corresponding temperature, *i.e.*, deuterium substitution at the hydroxyl site has minor effect on the reaction kinetics in the temperature range under investigation.

Fig. 4 plots the rate constants $k_3(T)$ measured over the temperature range 293 – 673 K for reaction (3) as well as $k_1(T)$ for reaction (1) for comparison purposes. The Arrhenius fit of $k_3$ data (in the units $cm^3$ $molecule^{-1}$ $s^{-1}$) yields

$$k_3(T) = (1.09 \pm 0.02) \times 10^{-11} \exp[-(640.00 \pm 7.23)/T], \qquad (10)$$

which gives an activation energy ~74% larger than that of reaction (1). It is apparent from Fig. 4 that the values of $k_3(T)$ are smaller than those of $k_1(T)$ at each corresponding temperature and $k_3(T)$ varies with temperature more dramatically than $k_1(T)$ does, strongly indicating that deuterium substitution from the methyl instead of hydroxyl site has pronounced effect on the reaction kinetics involved.

Now that the reaction kinetics is found to be overwhelmingly affected by the methyl rather than hydroxyl group in methanol, one would expect that the kinetics of the $C_2(a^3\Pi_u)$ reaction with the fully deuterated methanol, $CD_3OD$, i.e., reaction (4) may behave quite similarly to that of reaction (3). Plotted in Fig. 5 are the rate constants measured over the temperature range 293 – 673 K for reaction (4), $k_4(T)$. Fitting $k_4$ data (in the units $cm^3$ $molecule^{-1}$ $s^{-1}$) in the Arrhenius form gives

$$k_4(T) = (1.12 \pm 0.01) \times 10^{-11} \exp[-(666.37 \pm 4.63)/T]. \qquad (11)$$

Again, the positive temperature dependence of $k_4$ appears almost identical to that of $k_3$ which is also plotted in Fig. 5, as in the case for the comparison of $k_2$ with $k_1$ (Fig. 3) described above. In addition, $k_4(T) \approx k_3(T)$ at each corresponding temperature. All this further points to the fact that deuterium substitution at the hydroxyl site barely affects the reaction kinetics over the temperature range 293 – 673 K.

It is worth noting that although $k_1$ slightly differs from $k_2$ and the same holds for $k_3$ and $k_4$ (as can be discerned from Fig. 3 and 5, respectively), the sums $k_1 + k_4$ and $k_2 + k_3$ (compiled in Table 2) are found to be equal, for all temperatures between 293 and 673 K, to within an average deviation of ~0.3%, *i.e.*, the four temperature dependences sum to provide internally consistent results. The equality of

$$k_1 + k_4 = k_2 + k_3 \qquad (12)$$

suggests that the separation by reactive sites can work well for the reactions (1) – (4), *i.e.*, the following assumptions are rational:[45]

$$k_1 = k_{CH_3} + k_{OH}, \qquad (13)$$

$$k_2 = k_{CH_3} + k_{OD}, \qquad (14)$$

$$k_3 = k_{CD_3} + k_{OH}, \qquad (15)$$

and $\qquad k_4 = k_{CD_3} + k_{OD}, \qquad (16)$

hence $k_1/k_3$ can be thought of as reflecting principally $k_{CH_3}/k_{CD_3}$.[44]

The deuterium kinetic isotope effects, $k_1/k_2$ and $k_1/k_3$, display different temperature dependences that are plotted as $\log(k_1/k_2)$ and $\log(k_1/k_3)$ *versus* $1000/T$ in Fig. 6 and listed in Table 3, from which one can readily detect that $k_1/k_2$ keeps constant (~1.00) throughout the entire temperature range, whereas $k_1/k_3$ falls in a range of 1.8 – 3.1 and exhibits a clear negative temperature dependence with a best-fit expression of the form

$$k_1/k_3(T) = (1.22 \pm 0.03) \exp[(271.80 \pm 11.70)/T]. \qquad (17)$$

Such a negative temperature dependence is a straightforward outcome of the fact that the activation energy of reaction (1) is smaller than that of reaction (3) [*cf.* eqn (8) and (10)], which can be simply translated to that the C-H bond is more vulnerable to attack by $C_2(a^3\Pi_u)$ than the C-D bond.   This is explicitly correlated with BDE of the C-H and C-D bonds as it is well known that the former is smaller than the latter.[44]

The above analyses can be sorted out and summarized as follows: (i) Over the temperature range 293 – 673 K the observed deuterium kinetic isotope effects, $k_1(T) \approx k_2(T)$ and $k_3(T) \approx k_4(T)$, indicate conclusively that the hydroxyl group in methanol is an "uninvolved" reactive site and it, whether being deuterated or not, does not affect the reaction kinetics at its neighboring site, namely $CH_3$ or $CD_3$; (ii) The experimental observations of $k_3$, $k_4(T) < k_1, k_2(T)$ over the temperature range 293 – 673 K, together with (i), enable us to argue that the rate constants $k$ strongly correlate with BDE of the C-H(D) bonds because deuterium substitution occurring at the "involved" reactive methyl site dramatically slows down the reaction of $C_2(a^3\Pi_u)$ with methanol; (iii) Based on the experimental observations that all the rate constants $k(T)$ show a positive temperature dependence, together with (ii), a conclusion can be drawn that a site-specific hydrogen abstraction from the methyl group in methanol does indeed dominate reactivity over the temperature range under investigation, because if the reaction proceeds by $C_2(a^3\Pi_u)$ insertion into a C-H bond of the methanol one would expect that the reaction usually exhibits a negative temperature dependence and appears to show no kinetic isotope effect.[46-49]

## Conclusion

We have presented a first temperature dependence and kinetic isotope effect study on the reaction of $C_2(a^3\Pi_u)$ with methanol.   The bimolecular rate constants for a series of methanol isotopomers have been measured as a function of temperature between 293 and 673 K. The observed positive temperature dependences as well as deuterium kinetic isotope effects permit determination of the reaction mechanism as a site-specific hydrogen abstraction from the methyl rather than hydroxyl site.

## Acknowledgements

We gratefully acknowledge support from the National Natural Science Foundation of China (Grant Nos. 20673107 and 20873133), the Ministry of Science and Technology of China (Grant Nos. 2007CB815203 and 2010CB923302), the Chinese Academy of Sciences (KJCX2-YW-N24), and the Scientific Research Foundation for the Returned Overseas Chinese Scholars, Ministry of Education of China.

# Figure Captions

**Fig. 1** Typical time-resolved LIF decay profile for $C_2(a^3\Pi_u)$ in the presence of $C_2Cl_4$, $CH_3OD$, and excess Ar buffer gas with a total pressure of ~9.5 Torr at 573 K. $[C_2Cl_4] = 1.41 \times 10^{14}$ molecules cm$^{-3}$ and $[CH_3OD] = 1.42 \times 10^{15}$ molecules cm$^{-3}$. The solid curve is an exponential fit of the experimental data, which yields a pseudo-first-order rate constant $k'$.

**Fig. 2** Plots of pseudo-first-order rate constants *versus* the concentrations of $CH_3OH$, $CD_3OH$, $CH_3OD$, and $CD_3OD$ at 373, 373, 473 and 473 K, respectively, in the presence of excess Ar buffer gas with a total pressure of ~9.5 Torr. The solid lines are linear least-squares fits of the experimental data.

**Fig. 3** Arrhenius plots (displayed on a semi-logarithm scale) comparing the kinetic data for the reactions of $C_2(a^3\Pi_u)$ with $CH_3OH$ (filled square, $k_1$) and with $CH_3OD$ (filled downward triangle, $k_2$) over the temperature range 293 – 673 K. The error bars represent ±2σ estimates of the total experimental error. The solid lines are linear least-squares fits of the experimental data.

**Fig. 4**  Arrhenius plots (displayed on a semi-logarithm scale) comparing the kinetic data for the reactions of $C_2(a^3\Pi_u)$ with $CH_3OH$ (filled square, $k_1$) and with $CD_3OH$ (filled upward triangle, $k_3$) over the temperature range 293 – 673 K.  The error bars represent $\pm 2\sigma$ estimates of the total experimental error.  The solid lines are linear least-squares fits of the experimental data.

**Fig. 5**  Arrhenius plots (displayed on a semi-logarithm scale) comparing the kinetic data for the reactions of $C_2(a^3\Pi_u)$ with $CD_3OH$ (filled upward triangle, $k_3$) and $CD_3OD$ (filled circle, $k_4$) over the temperature range 293 – 673 K.  The error bars represent $\pm 2\sigma$ estimates of the total experimental error.  The solid lines are linear least-squares fits of the experimental data.

**Fig. 6**  Kinetic isotope effects, $k_1/k_2$ (open circle) and $k_1/k_3$ (open square), as a function of $1000/T(K)$.  The error bars represent $\pm 2\sigma$ estimates of the total experimental error.  The solid lines are linear least-squares fits (displayed on a semi-logarithm scale) of the data points.

**Fig. 1 (Hu *et al.*)**

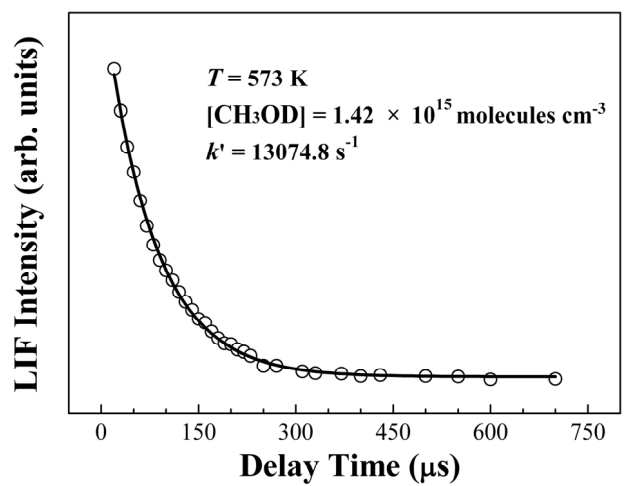

**Fig. 2 (Hu *et al.*)**

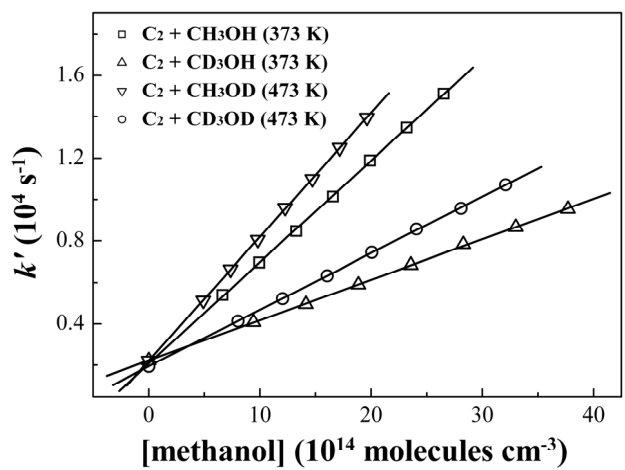

**Fig. 3 (Hu *et al.*)**

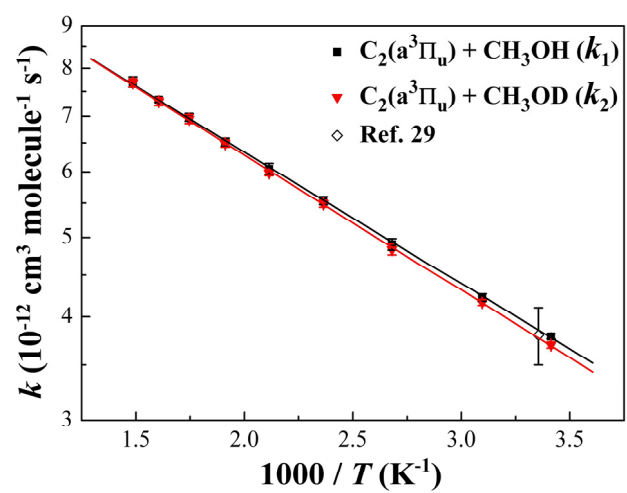

**Fig. 4  (Hu *et al.*)**

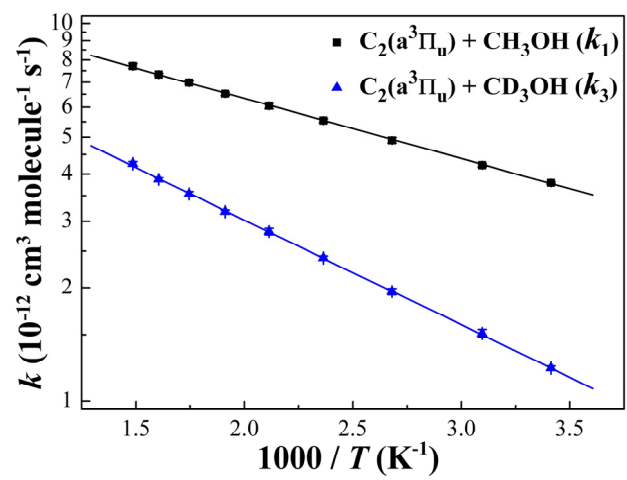

**Fig. 5 (Hu *et al.*)**

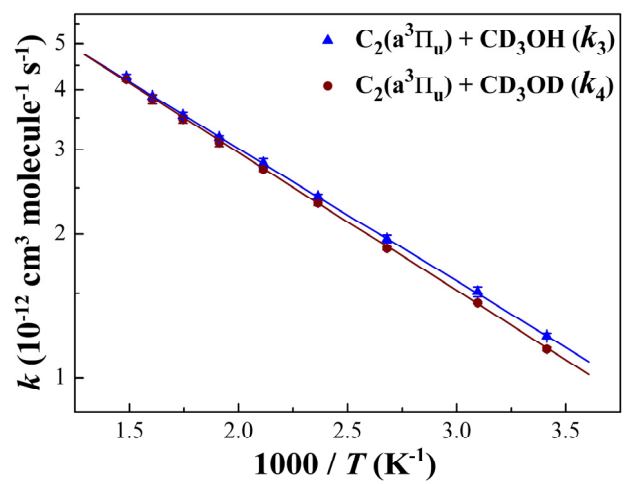

**Fig. 6  (Hu *et al.*)**

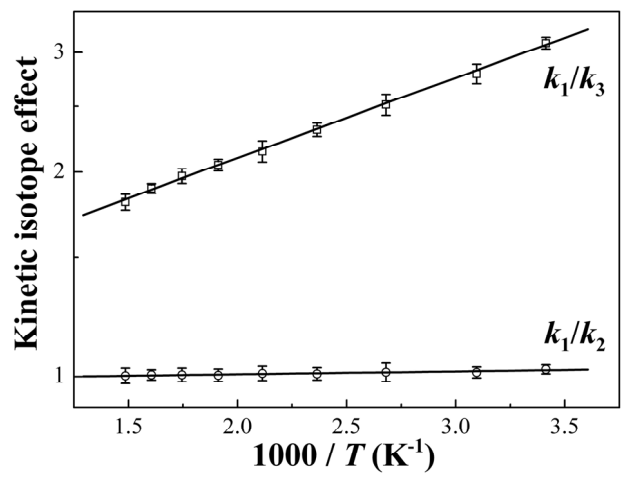

**Table 1** Bimolecular rate constants $k$ (in the units $10^{-12}$ cm$^3$ molecule$^{-1}$ s$^{-1}$) for reactions of $C_2(a^3\Pi_u)$ with CH$_3$OH ($k_1$), CH$_3$OD ($k_2$), CD$_3$OH ($k_3$), and CD$_3$OD ($k_4$) over the temperature range 293 – 673 K. The error bars represent ±2σ estimates of the total experimental error.

| $T$ (K) | $k_1$ | $k_2$ | $k_3$ | $k_4$ |
|---|---|---|---|---|
| 293 | 3.79 ± 0.03 | 3.70 ± 0.03 | 1.22 ± 0.02 | 1.15 ± 0.01 |
| 323 | 4.22 ± 0.04 | 4.16 ± 0.03 | 1.51 ± 0.04 | 1.43 ± 0.01 |
| 373 | 4.90 ± 0.08 | 4.84 ± 0.07 | 1.95 ± 0.04 | 1.86 ± 0.01 |
| 423 | 5.54 ± 0.06 | 5.49 ± 0.05 | 2.40 ± 0.03 | 2.32 ± 0.03 |
| 473 | 6.05 ± 0.10 | 5.98 ± 0.05 | 2.82 ± 0.06 | 2.73 ± 0.03 |
| 523 | 6.51 ± 0.07 | 6.49 ± 0.06 | 3.18 ± 0.03 | 3.11 ± 0.06 |
| 573 | 6.98 ± 0.08 | 6.95 ± 0.09 | 3.54 ± 0.05 | 3.48 ± 0.06 |
| 623 | 7.33 ± 0.05 | 7.29 ± 0.08 | 3.87 ± 0.04 | 3.82 ± 0.08 |
| 673 | 7.70 ± 0.11 | 7.68 ± 0.09 | 4.26 ± 0.05 | 4.21 ± 0.04 |

**Table 2** Sums of bimolecular rate constants $k$ (in the units $10^{-12}$ cm$^3$ molecule$^{-1}$ s$^{-1}$) for evaluating the equality stated in eqn (12). The uncertainties represent ±2σ estimates of the total experimental error; the $\sigma_{sum}$ values were obtained through propagation of errors using the $\sigma_k$ values listed in Table 1.

| $T$ (K) | $k_1 + k_4$ | $k_2 + k_3$ |
|---|---|---|
| 293 | 4.94 ± 0.03 | 4.92 ± 0.04 |
| 323 | 5.65 ± 0.05 | 5.67 ± 0.05 |
| 373 | 6.76 ± 0.08 | 6.79 ± 0.08 |
| 423 | 7.86 ± 0.07 | 7.89 ± 0.06 |
| 473 | 8.78 ± 0.10 | 8.78 ± 0.08 |
| 523 | 9.62 ± 0.09 | 9.67 ± 0.07 |
| 573 | 10.46 ± 0.10 | 10.49 ± 0.10 |
| 623 | 11.15 ± 0.09 | 11.16 ± 0.09 |
| 673 | 11.91 ± 0.12 | 11.94 ± 0.10 |

**Table 3** Deuterium kinetic isotope effects at the hydroxyl ($k_1/k_2$) and methyl ($k_1/k_3$) sites. The uncertainties represent ±2σ estimates of the total experimental error; the $\sigma_{sum}$ values were obtained through propagation of errors using the $\sigma_k$ values listed in Table 1.

| $T$ (K) | $k_1/k_2$ | $k_1/k_3$ |
| --- | --- | --- |
| 293 | 1.02 ± 0.02 | 3.10 ± 0.06 |
| 323 | 1.01 ± 0.02 | 2.79 ± 0.09 |
| 373 | 1.01 ± 0.03 | 2.51 ± 0.09 |
| 423 | 1.01 ± 0.02 | 2.31 ± 0.05 |
| 473 | 1.01 ± 0.02 | 2.15 ± 0.08 |
| 523 | 1.00 ± 0.02 | 2.05 ± 0.04 |
| 573 | 1.00 ± 0.02 | 1.97 ± 0.05 |
| 623 | 1.01 ± 0.02 | 1.89 ± 0.03 |
| 673 | 1.00 ± 0.03 | 1.81 ± 0.05 |